\documentclass{elsart5p}

\usepackage{graphics}
\usepackage{epsfig}

\journal{Physics Letters B}

\begin{document}

\renewcommand{\textfraction}{0.00000000001}
\renewcommand{\floatpagefraction}{1.0}
\begin{frontmatter}
\title{The isospin structure of photoproduction of $\pi\eta$ pairs from the nucleon in the threshold region}

\author[Basel]{A.~K{\"a}ser},   
\author[Mainz]{J.~Ahrens},             
\author[Glasgow]{J.R.M.~Annand},          
\author[Mainz]{H.J.~Arends},
\author[Kent]{K.~Bantawa},
\author[Mainz]{P.A.~Bartolome},            
\author[Bonn]{R.~Beck},         
\author[Petersburg]{V.~Bekrenev},
\author[Giessen]{H.~Bergh\"auser},            
\author[Pavia]{A.~Braghieri},           
\author[Edinburgh]{D.~Branford},            
\author[Washington]{W.J.~Briscoe},           
\author[UCLA]{J.~Brudvik},             
\author[Lebedev]{S.~Cherepnya},  
\author[Pavia]{S. Costanza},         
\author[Washington]{B.~Demissie},
\author[Basel]{M.~Dieterle},             
\author[Mainz,Glasgow,Washington]{E.J.~Downie},            
\author[Giessen]{P.~Drexler}, 
\author[Lebedev]{L.V.~Fil'kov},
\author[Tomsk]{A.~Fix},  
\author[Glasgow]{D.I.~Glazier}, 
\author[Glasgow]{D. Hamilton},          
\author[Mainz]{E.~Heid}, 
\author[Sackville]{D.~Hornidge}, 
\author[Glasgow]{D.~Howdle}, 
\author[Regina]{G.M.~Huber},           
\author[Mainz]{O.~Jahn},
\author[Basel]{I.~Jaegle},
\author[Edinburgh]{T.C.~Jude},
\author[Lebedev,Mainz]{V.L.~Kashevarov},
\author[Basel]{I.~Keshelashvili},        
\author[Moskau]{R.~Kondratiev},          
\author[Zagreb]{M.~Korolija},  
\author[Petersburg]{S.P.~Kruglov}, 
\author[Basel]{B.~Krusche}\ead{Bernd.Krusche@unibas.ch},  
\author[Moskau]{V.~Lisin},               
\author[Glasgow]{K.~Livingston},          
\author[Glasgow]{I.J.D.~MacGregor},       
\author[Basel]{Y.~Maghrbi},
\author[Glasgow]{J.~Mancell},  
\author[Kent]{D.M.~Manley}, 
\author[Washington]{Z.~Marinides},           
\author[Glasgow]{J.C.~McGeorge}, 
\author[Glasgow]{E.~McNicoll},          
\author[Zagreb]{D.~Mekterovic},          
\author[Giessen]{V.~Metag},
\author[Zagreb]{S.~Micanovic},
\author[Sackville]{D.G.~Middleton},
\author[Pavia]{A.~Mushkarenkov},               
\author[Bonn]{A.~Nikolaev},   
\author[Giessen]{R.~Novotny},
\author[Basel]{M.~Oberle},
\author[Mainz]{M.~Ostrick},
\author[Mainz]{P.~Otte}
\author[Mainz,Washington]{B.~Oussena},
\author[Pavia]{P.~Pedroni}, 
\author[Basel]{F.~Pheron},            
\author[Moskau]{A.~Polonski},            
\author[UCLA]{S.N.~Prakhov}, 
\author[Glasgow]{J.~Robinson},                    
\author[Glasgow]{G.~Rosner},              
\author[Basel,Pavia]{T.~Rostomyan},           
\author[Mainz]{S.~Schumann},
\author[Edinburgh]{M.H.~Sikora},            
\author[Catholic]{D.I.~Sober},               
\author[UCLA]{A.~Starostin},           
\author[Zagreb]{I.~Supek},               
\author[Mainz,Giessen]{M.~Thiel},        
\author[Mainz]{A.~Thomas},              
\author[Mainz,Bonn]{M.~Unverzagt},                       
\author[Edinburgh]{D.P.~Watts},
\author[Basel]{D.~Werthm\"uller\thanksref{1}},
\author[Basel]{L.~Witthauer} 

\address[Basel] {Department of Physics, University of Basel, Ch-4056 Basel, Switzerland}
\address[Tomsk] {Laboratory of Mathematical Physics, Tomsk Polytechnic University, Tomsk, Russia}
\address[Mainz] {Institut f\"ur Kernphysik, University of Mainz, D-55099 Mainz, Germany}
\address[Glasgow] {SUPA School of Physics and Astronomy, University of Glasgow, G12 8QQ, United Kingdom}
\address[Kent] {Kent State University, Kent, Ohio 44242, USA}
\address[Bonn] {Helmholtz-Institut f\"ur Strahlen- und Kernphysik, University Bonn, D-53115 Bonn, Germany}  
\address[Petersburg] {Petersburg Nuclear Physics Institute, RU-188300 Gatchina, Russia}
\address[Giessen] {II. Physikalisches Institut, University of Giessen, D-35392 Giessen, Germany}
\address[Pavia] {INFN Sezione di Pavia, I-27100 Pavia, Pavia, Italy}
\address[Edinburgh] {SUPA School of Physics, University of Edinburgh, Edinburgh EH9 3JZ, United Kingdom}
\address[Washington] {Center for Nuclear Studies, The George Washington University, Washington, DC 20052, USA}
\address[UCLA] {University of California Los Angeles, Los Angeles, California 90095-1547, USA}
\address[Lebedev] {Lebedev Physical Institute, RU-119991 Moscow, Russia}
\address[Sackville] {Mount Allison University, Sackville, New Brunswick E4L1E6, Canada}
\address[Regina] {University of Regina, Regina, SK S4S-0A2 Canada}
\address[Moskau] {Institute for Nuclear Research, RU-125047 Moscow, Russia}
\address[Zagreb] {Rudjer Boskovic Institute, HR-10000 Zagreb, Croatia}
\address[Catholic] {The Catholic University of America, Washington, DC 20064, USA}

\thanks[1] {present address: SUPA School of Physics and Astronomy, University of Glasgow, G12 8QQ, United Kingdom}

\begin{abstract}
Photoproduction of $\pi\eta$-pairs from nucleons has been investigated from 
threshold up to incident photon energies of $\approx$~1.4~GeV.
The quasi-free reactions $\gamma p\rightarrow p\pi^0\eta$, 
$\gamma n\rightarrow n\pi^0\eta$, $\gamma p\rightarrow n\pi^+\eta$,
and $\gamma n\rightarrow p\pi^-\eta$ were for the first time measured from 
nucleons bound in the deuteron. The corresponding reactions from a free-proton 
target were also studied to investigate final-state interaction effects 
(for neutral pions the free-proton results could be compared to previous 
measurements; the $\gamma p\rightarrow n\pi^+\eta$ reaction was measured 
for the first time). For the $\pi^0\eta$ final state coherent production 
via the $\gamma d\rightarrow d\pi^0\eta$ reaction was also investigated.
The experiments were performed at the tagged photon beam of the 
Mainz MAMI accelerator using an almost $4\pi$ coverage electromagnetic 
calorimeter composed of the Crystal Ball and TAPS detectors. 
The total cross sections for the four different final states obey
the relation $\sigma(p\pi^0\eta)$ $\approx$ $\sigma(n\pi^0\eta)$ $\approx$ 
$2\sigma(p\pi^-\eta)$ $\approx$ $2\sigma(n\pi^+\eta)$ as expected
for a dominant contribution from a 
$\Delta^{\star}\rightarrow\eta\Delta(1232)\rightarrow\pi\eta N$ reaction chain,
which is also supported by the shapes of the invariant-mass distributions of 
nucleon-meson and $\pi$-$\eta$ pairs. The experimental results are compared to the 
predictions from an isobar reaction model.
  \end{abstract}
\end{frontmatter}

\section{Introduction}
Photoproduction of mesons is a well established tool for the investigation of excited
states of the nucleon. Reactions with meson pairs in the 
final state have gained a lot of interest because they allow the study of resonances 
that have small decay branching ratios to the nucleon ground state and decay instead
via intermediate excited states. 

The best studied double-meson final state is pion pairs. In particular, double 
$\pi^0$ production has been studied 
(see e.g. \cite{Sarantsev_08,Thoma_08,Krambrich_09,Kashevarov_12,Zehr_12,Oberle_13}).
This reaction has the advantage that due to the small coupling of photons to neutral pions 
non-resonant background contributions are small. However, recently the $\pi\eta$ 
final state has also attracted interest. Total cross sections, invariant mass distributions, 
and also some polarization observables have been measured for the 
$\gamma p\rightarrow p\pi^0\eta$  reaction at LNS in Sendai, Japan \cite{Nakabayashi_06}, 
GRAAL at ESRF in Grenoble, France \cite{Ajaka_08}, ELSA in Bonn, Germany 
\cite{Horn_08a,Horn_08b,Gutz_08,Gutz_10,Gutz_14}, and at MAMI in Mainz, Germany 
\cite{Kashevarov_09,Kashevarov_10} (see \cite{Krusche_15} for a recent summary). 
This decay channel is more selective than double-$\pi^0$ production. The $\eta$-meson 
is isoscalar, such that nucleon resonances can only emit it in 
$N^{\star}\rightarrow N^{(\star )}$ or $\Delta^{\star}\rightarrow \Delta^{(\star )}$ 
transitions. Thus, one expects that two classes of nucleon resonances are 
important for this reaction: excited $\Delta^{\star}$-states with significant 
$\eta$-decays to the $\Delta$(1232) and $N^{\star}$ resonances decaying to 
$N(1535)1/2^-$ via pion emission. The main signature of the first decay type are 
pion - nucleon invariant masses peaking at the $\Delta$(1232) mass, while the latter 
will produce $\eta$ - nucleon invariant masses close to the $N(1535)1/2^-$ position. 
These two components have been identified in the previously measured invariant mass 
distributions, which in addition show a signal from the $a_{0}(980)\rightarrow\pi\eta$ 
decay in the $\eta$ - $\pi$  invariant mass \cite{Horn_08a,Horn_08b,Gutz_14}. 

These analyses discovered a strong dominance of the 
$\Delta 3/2^-\rightarrow \eta\Delta\mbox{(1232)}\rightarrow\pi^0\eta p$
decay chain; in the threshold region from the $\Delta(1700)3/2^-$ and at higher
incident photon energies from the $\Delta(1940)3/2^-$ 
\cite{Horn_08b,Kashevarov_09,Fix_10}. So far, there were only two cases where
a photon-induced meson-production reaction is completely dominated by a single
resonance and allows its almost background free study: pion production in the
range of the $\Delta$ resonance and $\eta$ production via the $N(1535)1/2^-$ 
resonance \cite{Krusche_95,Krusche_97}. Photoproduction of $\pi\eta$-pairs 
offers the same chance for the $\Delta(1700)3/2^-$, and in fact it has 
already been used in \cite{Gutz_14,Kashevarov_09} to extract parameters of this 
state. This is important because the structure of the $\Delta(1700)3/2^-$ is still 
under discussion.
D\"oring, Oset, and Strottman \cite{Doering_06a,Doering_06b} have studied this resonance
with a coupled-channel chiral unitary approach for meson-baryon scattering in which it 
is dynamically generated. They also predict a dominant contribution of  
$\Delta 3/2^-\rightarrow \eta\Delta\mbox{(1232)}\rightarrow\pi\eta N$ to the 
$\pi\eta N$ final state. 

This decay chain is characterized by its spin and isospin structure, 
which is much different from single $\eta$ production in the threshold region. 
The $\Delta(1700)3/2^-$ state can be electromagnetically excited by
electric dipole ($E1$) or magnetic quadrupole ($M2$) photons. 
The $\Delta 3/2^-\rightarrow \Delta 3/2^+\eta$ decay is only possible for the $\eta$
emitted in relative $s$- or $d$-wave ($L_{\eta}$=0,2), and the $s$-wave is expected
to dominate in the threshold region. The pion from the $\Delta 3/2^+\rightarrow N\pi$ decay
is emitted in a relative $p$-wave ($L_{\pi}$=1). Therefore, the reaction may involve 
spin-flip and non-spin-flip transitions. Their relative strengths can be calculated 
\cite{Egorov_13} from the ratio of the helicity couplings $a=A_{3/2}/A_{1/2}$ of 
the $\Delta 3/2^-$ resonance. The most recent values from the Particle Data Review 
\cite{PDG} for the helicity couplings result in $a\approx 1$, which corresponds to
$\sigma_K /\sigma_L\approx 0.6$ \cite{Egorov_13}, where $\sigma_K$ and $\sigma_L$
are the spin-flip and spin-independent components of the reaction, respectively. 
The large contribution of the spin-independent part is in sharp contrast to single-$\eta$ 
production via the dominating $N(1535)1/2^-$ resonance, which proceeds only through 
the spin-flip term. This has important consequences for the coherent production of
$\pi^0\eta$ pairs from nuclei as compared to coherent single $\eta$ production,
which is forbidden for spin $J=0$ nuclei. Coherent $\eta$ production has been 
investigated as a possible doorway for the formation of $\eta$-mesic states
\cite{Pfeiffer_04,Pheron_12,Maghrbi_13}, but cannot be used for some of the most
promising candidates like $^4$He nuclei. Production of $\pi^0\eta$ pairs can
avoid this problem.

The situation for the isospin structure is even more simple. The electromagnetic 
excitation of the $\Delta$-resonance is identical for protons and neutrons. 
From the isoscalar nature of the $\eta$ ($I_{\eta}$=0) and the isovector nature 
of the pion ($I_{\pi}$=1), it follows immediately that the sequential reaction chain 
from the $\Delta 3/2^-$ via the $\Delta 3/2^+\eta$ intermediate state must 
have the same isospin pattern as single photoproduction of pions through the 
$\Delta$-resonance \cite{Krusche_03}. Thus, applying Clebsch-Gordon
coefficients gives
\begin{eqnarray} 
 \sigma (\gamma p\rightarrow \eta\pi^0 p) & = &
 \sigma (\gamma n\rightarrow \eta\pi^0 n) =\nonumber\\
2\sigma (\gamma p\rightarrow \eta\pi^+ n) & = &
2\sigma (\gamma n\rightarrow \eta\pi^- p)\;.
\label{eq:isorel}
\end{eqnarray}
A test of these relations would give additional weight to the proposed dominance of
the $\Delta$-resonance decay. Any deviations would point to significant contributions
from other resonances or non-resonant backgrounds. So far, only data for the 
$\gamma p\rightarrow \pi^0\eta p$ reaction is available. Also, the isospin structure
is very favorable for coherent production of $\pi^0\eta$ pairs. Since the electromagnetic
helicity couplings for $\Delta$-resonances are identical for protons and neutrons,
no cancellations can occur. 

The present paper summarizes the results from the measurement of quasi-free production
of $\pi\eta$ pairs from nucleons bound in the deuteron for all four possible isospin 
combinations from Eq.~\ref{eq:isorel} up to incident photon energies of 1.4 GeV. 
Total cross sections have been extracted for all four reaction channels and are compared 
to the results from the reaction model discussed in \cite{Kashevarov_09}. In addition, 
the total cross section for the coherent reaction $\gamma d\rightarrow d\pi^0\eta$ has 
been determined and compared to the model results from Egorov and Fix \cite{Egorov_13}. 

\section{Experiment and Analysis}
\label{sec:setup}

The experiment was carried out at the Mainz MAMI accelerator \cite{Herminghaus_83,Walcher_90}
using a quasi-monochromatic photon beam with energies up to $\approx$1.4 GeV from
the Glasgow tagged photon spectrometer \cite{Anthony_91,Hall_96,McGeorge_08}. The
primary electron beam (energy of 1.508 GeV, for some part of the beam time 1.577 MeV see \cite{Oberle_14})
produced bremsstrahlung photons in a copper radiator of 10 $\mu$m thickness. The energy resolution of
the photon beam is related to the 4 MeV bin width of the tagger focal plane 
detectors. 

Three different beam times with liquid deuterium targets and one
measurement with a liquid hydrogen target for control were analyzed for the results 
summarized in this work. The same data set has already been analyzed in 
\cite{Oberle_13,Oberle_14} for beam-helicity asymmetries in quasi-free photoproduction
of $\pi^0\pi^0$ and $\pi^0\pi^{\pm}$ pairs, for total cross sections and angular
distributions in single $\eta$ production \cite{Werthmueller_14} and single $\pi^0$
production \cite{Dieterle_14}. Details for the setup, target and beam parameters, and 
analysis procedures are discussed in these references. Here, only a short summary
is given. 

The detector setup (see \cite{Oberle_14,Werthmueller_14} for schematic drawings
and details) combined the Crystal Ball (CB) \cite{Starostin_01} and TAPS 
\cite{Novotny_91,Gabler_94} electromagnetic calorimeters (CB: 672 NaI(Tl) crystals, 
TAPS: 384 BaF$_2$ crystals). The targets were mounted in the center of the CB and 
surrounded by a detector for charged particle identification (PID) \cite{Watts_04}. 
For the same purpose, all TAPS modules had individual plastic scintillators in front 
of the crystals (`TAPS-Veto'). The combined setup covered $\approx$~97\% of the full 
solid angle and detected photons from the decays of the neutral mesons, charged pions, 
protons, neutrons, and deuterons.

\begin{figure}[thb]
\centerline{\resizebox{0.5\textwidth}{!}{
  \epsfig{file=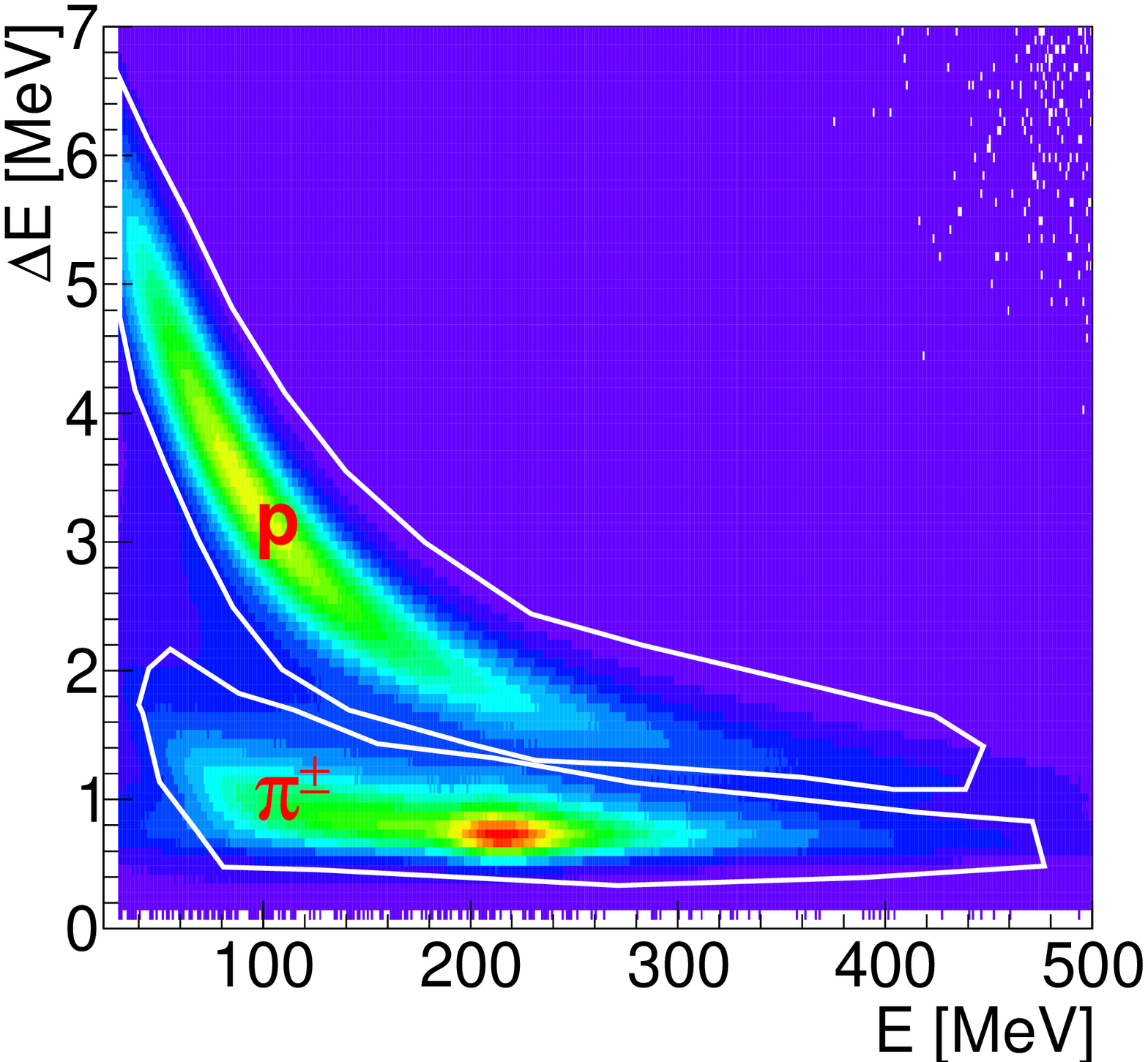,scale=0.5}
  \epsfig{file=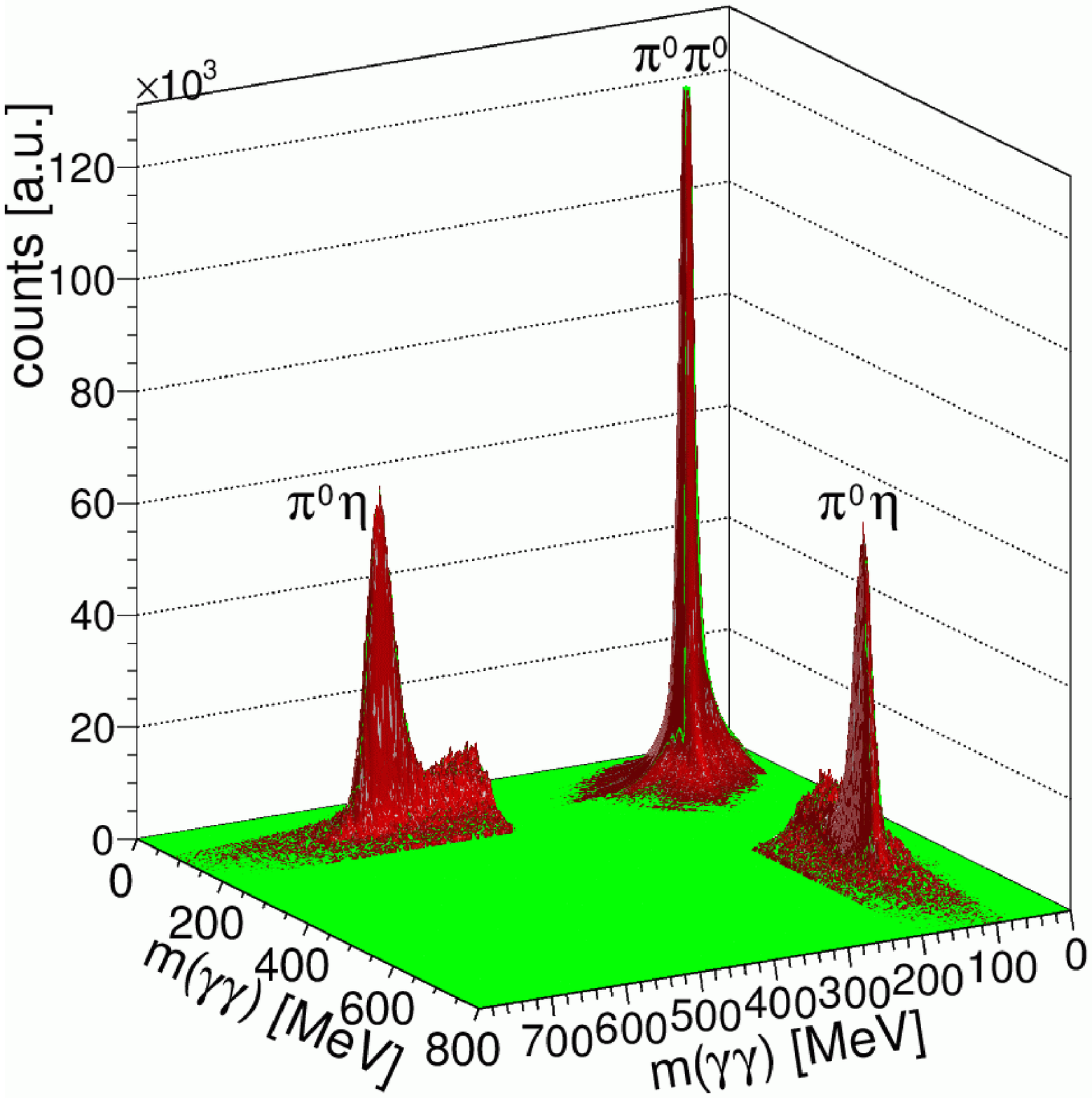,scale=0.45}
}}
\caption{Left hand side: Identification of protons and charged pions in CB with 
a $\Delta E-E$ analysis. White lines indicate the accepted areas. 
Right hand side: Two-dimensional invariant-mass distribution for events with four photons. 
The (red) regions around the $\pi^0\eta$ peaks are scaled up by a factor of 50.} 
\label{fig:invmas} 
\end{figure}

The analysis for the quasi-free reactions was almost identical to that for the
$\pi^0\pi^0$ and $\pi^0\pi^{\pm}$ final states discussed in \cite{Oberle_13} and
in \cite{Oberle_14}. In the first step, hits in both calorimeters were classified as 
`charged' or `neutral' depending on the response of the PID and the TAPS-Veto. 
Charged hits in CB were separated into protons and charged pions using the $\Delta E-E$ 
analysis of the PID and CB. The result is shown in Fig.~\ref{fig:invmas}. 
No direct separation between photons and neutrons was possible for hits in the CB. 
Neutral hits in the CB were therefore, in this stage of the analysis, accepted as 
candidates for photons and neutrons.

For hits in TAPS, photons and neutrons were separated as in 
\cite{Oberle_13,Oberle_14,Werthmueller_14} using a pulse-shape analysis (PSA) for 
the BaF$_2$ signals and a time-of-flight (ToF) versus energy analysis. Protons in 
TAPS were also required to have passed the PSA as nucleons, and the ToF-versus-E 
as protons. Recoil deuterons in TAPS were separated (for events with a $\pi^0\eta$ pair) 
from protons in the ToF-versus-E spectra, where they appeared as a clearly 
separated band. Charged pions in TAPS were not included in the analysis because the 
only identification possibility would have been ToF-versus-E, but protons 
(partly from reactions with much higher cross sections) that leaked into the pion band 
produced substantial background. This means that for the $\pi^+\eta$ and $\pi^-\eta$ 
final states, a small part of the total reaction phase space (charged pions
at laboratory polar angles less than 20$^{\circ}$) was not in the acceptance of the 
detector. 

\begin{table}[h]
\begin{center}
\caption{Selected event classes for the cross sections $\sigma_p$ (coincident with 
recoil protons), $\sigma_n$ (coincident with recoil neutrons), $\sigma_{\rm incl}$ 
(no condition for recoil nucleons), $\sigma_d$ (coincident with recoil deuterons) 
for $\pi\eta$-pairs with neutral and charged pions. $n$ and $c$ mark neutral and 
charged hits in the calorimeter, respectively (distinguished by the response of 
the charged-particle detectors).
} 
\label{tab:events}       
\begin{tabular}{|c|c|c|c|c|}
\hline\noalign{\smallskip}
& $\sigma_p$ & $\sigma_n$ & $\sigma_{\rm incl}$ & $\sigma_d$\\
\hline
$\pi^0\eta$     & 4$n$ \& 1$c$ & 5$n$  & 4$n$ or 5$n$ or (4$n$ \& 1$c$) & 4n \& 1c\\
$\pi^{\pm}\eta$ & 2$n$ \& 2$c$ & 3$n$ \& 1$c$ & (2$n$ \& 1$c$) or (2$n$ \& 2$c$) or (3$n$ \& 1$c$) & -\\
\hline
\end{tabular}
\end{center}
\end{table}

\begin{figure*}[thb]
\centerline{\resizebox{0.9\textwidth}{!}{
  \epsfig{file=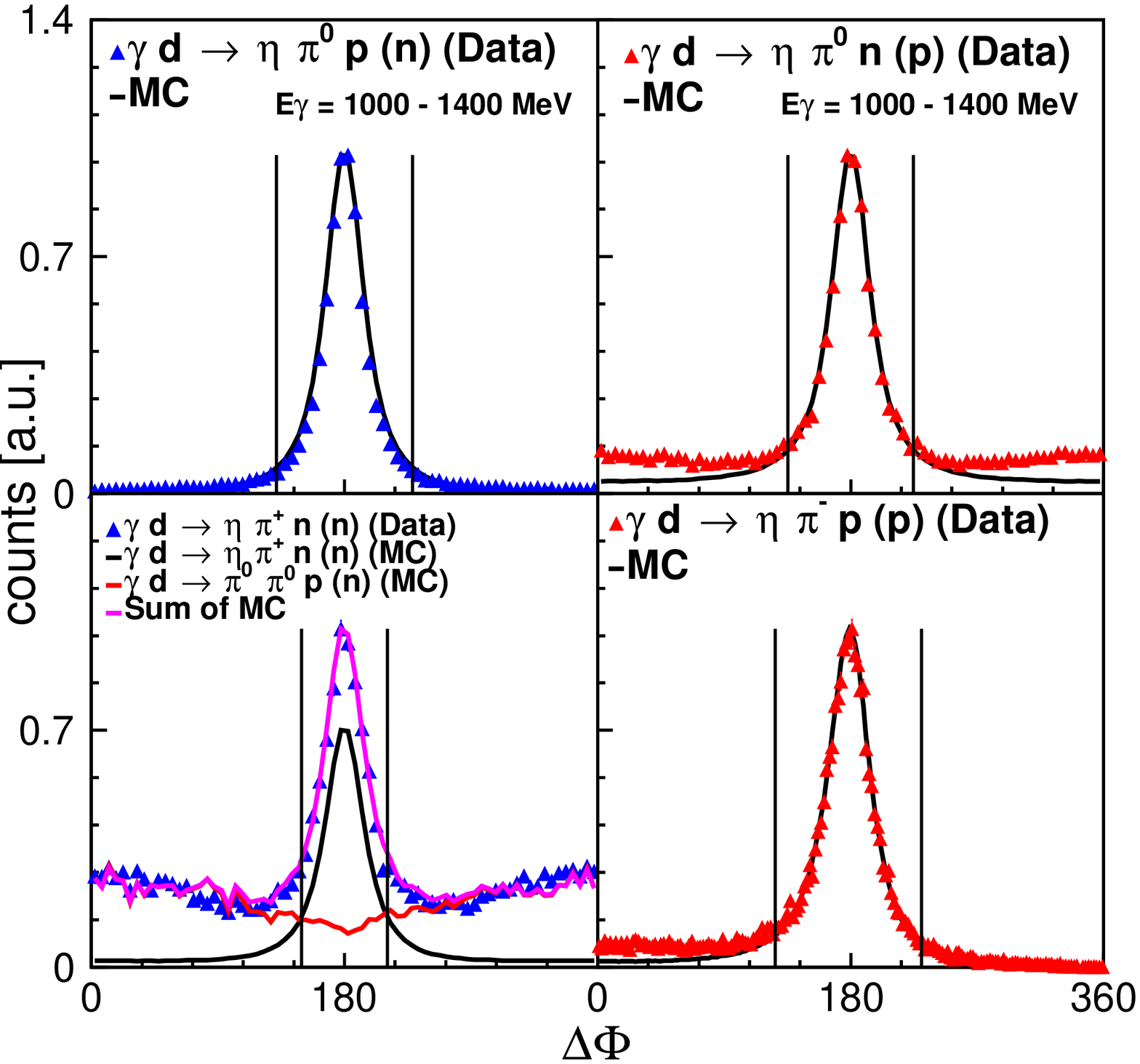,scale=0.45}
  \epsfig{file=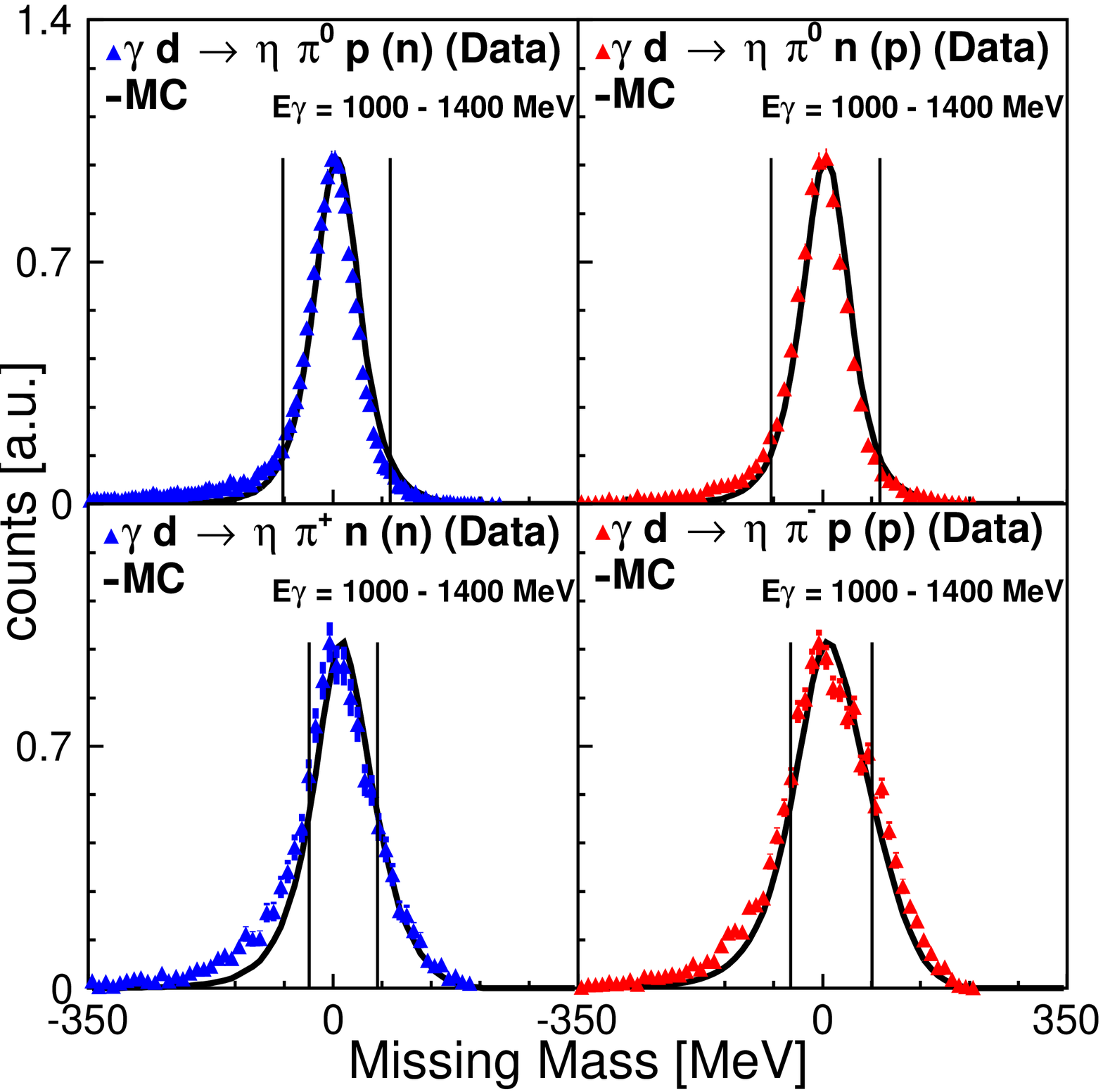,scale=0.42}
}}
\caption{Left hand side: Coplanarity spectra (see text) for the four different isospin channels 
(final state nucleons in brackets are spectators, without brackets are detected participants). 
Black curves: MC simulations of signal reactions. Red and magenta curves (only for
$\pi^+n\eta(n)$ final state, bottom, left corner): simulated background 
from $\pi^0\pi^0 p(n)$ final state. Vertical lines indicate applied cuts.
Right hand side: Missing mass spectra for same reactions.}
\label{fig:mismas} 
\end{figure*}

Seven different event classes were analyzed in total (see Table \ref{tab:events}). 
They correspond to both types of pions ($\pi^0$ or $\pi^{\pm}$),
to reactions with coincident recoil protons ($\sigma_p$) and coincident recoil neutrons
($\sigma_n$), the inclusive reaction $\sigma_{\rm incl}$ without any condition
for recoil nucleons, and for $\pi^0\eta$ in addition to the coherent
reaction (coincident with recoil deuterons). The indices $p$ and $n$ refer to the 
final-state nucleons, protons and neutrons respectively. The inclusive reaction 
$\sigma_{\rm incl}$ is independent from the recoil nucleon detection efficiencies 
and was used to check for systematic effects because it must satisfy the condition 
$\sigma_{\rm incl}\approx\sigma_p +\sigma_n+(\sigma_d)$ 
($\sigma_d$ contributes only for $\pi^0\eta$-pairs). 

In the second step of the analysis, a $\chi^2$ test of the invariant masses of pairs 
of neutral hits was made for events with three and more neutral hits, testing the 
hypothesis of $\pi^0$ and $\eta$ invariant masses. The $\chi^2$ of all possible
combinations of neutral hits to disjunct pairs was calculated from 
\begin{equation}\label{eq:chi2}
\chi^{2}(k) = \sum_{i=1}^{n_m}\left 
(\frac{m_{\pi^0,\eta}-m_{i,k}}{\Delta m_{i,k}}\right)^{2}~~~~~~~~~k=1,..,n_p,
\end{equation} 
where $n_m$ is the number of neutral mesons ($n_m$ = 1 for $\pi^{\pm}\eta$ final 
states, $n_m$ = 2 for $\pi^0\eta$) and $n_p$ is the number of different combinations 
of the neutral hits to pairs ($n_p$ = 1 for $2n$ events, $n_p$ = 15 for $5n$ events). 
The $m_{i,k}$ are the invariant masses of 
the i-th pair in the k-th permutation of the hits and the $\Delta m_{i,k}$ are 
the corresponding uncertainties computed event-by-event from the experimental
energy and angular resolution. The mesons nominal masses $m_{\pi^0,\eta}$ were chosen 
such that for each event with four or five neutral hits, the hypotheses of a 
$\pi^0\eta$ and a $\pi^0\pi^0$ pair and for events with two or three neutral hits, 
the $\eta$ and $\pi^0$ hypotheses were tested. Only the combination with the minimum 
$\chi^2$ was selected for further analysis. For events with an odd number of neutral 
hits, for which no hit was directly identified as a neutron, the one which was not 
identified as a meson decay photon by the $\chi^2$-test was assigned to the neutron.

The two-dimensional invariant mass spectrum for the $\pi^0\eta$ final state is shown 
in Fig.~\ref{fig:invmas}. The regions around the $\pi^0\eta$ peaks are scaled up by 
a factor of 50 with respect to the $\pi^0\pi^0$ peak. The remaining background below 
the peaks appears mainly as a band at the $\pi^0$ invariant mass running parallel
to the axis of $\eta$ invariant mass. The spectra were therefore projected onto
$\eta$-invariant masses. They were fitted with a third degree polynomial and the 
simulated line shape of the $\eta$ invariant mass peaks. Residual background was
removed in two further steps. The coplanarity of the events (i.e. the condition 
that in the photon-nucleon center-of-momentum (cm) system, the azimuthal angle between 
the recoil nucleon momentum and the sum of the meson momenta must be 180$^{\circ}$)
was tested. The result is shown in Fig.~\ref{fig:mismas}. Background for detection of 
$\pi^0\eta p$ triples is almost negligible, for $\pi^0\eta n$ triples it is small. 
In the final step, the missing mass of the events (although the recoil nucleon was 
detected, it was treated as a `missing' particle) was analyzed as described in 
\cite{Oberle_13} for double $\pi^0$ production. The resulting spectra (Fig.~\ref{fig:mismas}) 
were almost background free and the selection criteria indicated in the figure were applied. 

\begin{figure}[htb]
{\resizebox{0.5\textwidth}{!}{
  \epsfig{file=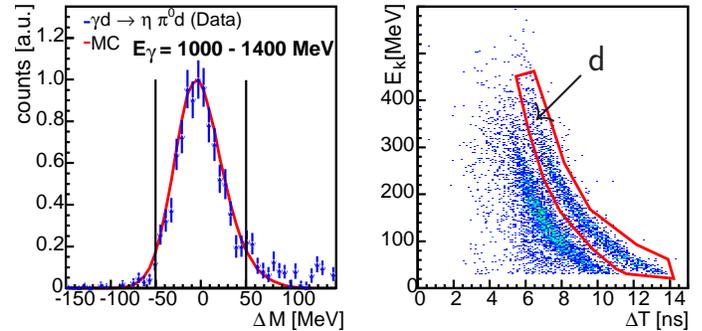,scale=0.5}
}}
\caption{Identification of $\gamma d\rightarrow d\pi^0\eta$. Left hand side: missing mass
spectrum, vertical lines indicate event selection (only events with missing mass in this range
are included in the ToF-versus-E spectrum), (red) curve: Monte Carlo simulation
of line shape. Right hand side: ToF-versus-E in TAPS, (red) lines indicate event selection
(only events with charged particle in this region included in missing mass spectrum).} 
\label{fig:coh_iden} 
\end{figure}

The identification of the coherent $\gamma d\rightarrow d\pi^0\eta$ reaction used the same 
analysis steps for the two neutral mesons. The recoil deuterons were identified in 
TAPS by a ToF-versus-E analysis. The final step was a missing-mass analysis (the deuteron 
was treated as missing particle). Typical spectra are shown in Fig.~\ref{fig:coh_iden}. 
The ToF-versus-E spectrum shows a clear deuteron band structure. The missing-mass 
spectrum is almost background free and perfectly reproduced by the Monte Carlo simulation 
(the signal width is much more narrow than for the quasi-free reactions). Also deuterons which were 
stopped in the TAPS-Veto were accepted. In this case, the ToF-versus-E analysis
was based on the signals from the TAPS-Veto detectors (not shown in the figure).    

The analysis of the $\pi^{\pm}\eta$ final state was carried out analogously to the 
analysis of $\pi^0\pi^{\pm}$ pairs in \cite{Oberle_14}. Charged pions were
selected by the $E-\Delta E$ cut shown in Fig.~\ref{fig:invmas} and the $\eta$ invariant 
mass peaks were fitted with background polynomials and simulated peak line shapes.
This was followed by coplanarity and missing-mass analysis cuts. The results are 
shown at the bottom of Fig.~\ref{fig:mismas}. The only significant background observed was 
in the coplanarity spectra of the $\pi^+\eta n$ final state. It is related to events from
the $\gamma d\rightarrow \pi^0\pi^0 p(n)$ reaction, for which one $\pi^0$-decay photon 
has escaped detection, the other photon from the same $\pi^0$ was misassigned as a 
neutron, and the proton was misidentified as a charged pion. The shape of this background 
could be reproduced by Monte Carlo simulations with GEANT4. The background 
below the 180$^{\circ}$ peak was subtracted and this was done individually for each data
point in the subsequent missing mass spectra. The resolution in the missing-mass spectra
is worse than for neutral pions because of the less precise measurement of the energies 
of the charged pion in the calorimeter. However, there is basically no background
visible in the missing mass spectra and the line shapes are well reproduced by
Monte Carlo simulations. 

Although Figs.~\ref{fig:invmas},\ref{fig:mismas} show spectra obtained by integration over 
all incident photon energies and other variables (angles, invariant masses) the actual analysis was 
performed individually for each bin of photon energy and other variables for differential
spectra. 

The extracted yields were converted to cross sections using the target surface densities
(0.231$\pm$0.005 barn$^{-1}$ for most of the beam times, see \cite{Oberle_14} for details),
the incident photon flux, the simulated detection efficiency, and the decay branching
ratios \cite{PDG} of $\pi^0$ (98.823$\pm$0.034\%) and $\eta$ (39.41$\pm$ 0.20 \%) mesons 
into photon pairs. More details are given in \cite{Werthmueller_14}. The detection
efficiency was simulated similarly as in \cite{Oberle_14,Werthmueller_14} with the GEANT4 
code \cite{GEANT4}. Since the reaction is dominated by the 
$\Delta 3/2^-\rightarrow \eta\Delta\mbox{(1232)}\rightarrow \pi\eta N$ chain, 
the sequential decay with the intermediate $\eta\Delta \mbox{(1232)}$ state was used
for the event generator. This generator describes the missing mass
distributions (see Fig.~\ref{fig:mismas} and the $\pi\eta$, $\pi N$, and $\eta N$ invariant 
mass distributions (not shown) very well.
The angular distributions deviate not much from isotropy, which was used for the MC 
(see \cite{Kashevarov_09} for angular distributions of the free $\gamma p\rightarrow p\pi^0\eta$ 
reactions, the results for the other isospin channels are similar).
Therefore, the simulation reflects all relevant properties of the reaction. 
As in \cite{Werthmueller_14}, corrections for the recoil nucleon
detection efficiencies obtained from direct measurements have been applied.

Total cross sections for final states with detected recoil nucleons were analyzed 
in two ways (see \cite{Werthmueller_14}
for details). The reconstruction as a function of the incident photon energy profits 
from the good energy resolution of the spectrometer, but the results are folded in 
with nuclear Fermi motion. 
In the second analysis, the Fermi-momentum corrected cm energy $W$ was reconstructed from the
kinematics of the final state. This eliminates the Fermi smearing, but introduces effects from 
experimental resolution (energies and angles measured with the calorimeters) into the 
$W$ measurement. However, for the present measurement no rapid 
variations of the cross sections are involved. Both effects are therefore 
not important. The quasi-free cross section data extracted by the two methods are in
good agreement (apart from the immediate threshold region where Fermi motion smears out 
the excitation function across the free production threshold). Furthermore, an analysis 
of free proton data with the $W$ reconstruction method (see Fig.~\ref{fig:total}c) gives 
an identical result as the use of the incident photon energy (in the latter case neither 
Fermi motion nor reconstruction resolution enter). 

Systematic uncertainties of the data have been discussed in detail
in \cite{Oberle_14,Werthmueller_14}. The total overall normalization uncertainty (photon
flux, target density) is between 5\% (quadratic addition) and 7\% (linear addition).
Uncertainties from analysis cuts and simulation of the detection efficiency excluding
the recoil nucleons are in the range of 5 - 10\%. The uncertainty from the recoil nucleon
detection efficiency has been estimated in \cite{Werthmueller_14} at the 10\% level.
It can be additionally checked by the comparison of the inclusive cross sections with
the sum of the exclusive ones (see Sec.~\ref{sec:results}). The overall normalization 
uncertainties cancel in all ratios. The effects from the meson detection cancel almost
completely for the $\sigma (\pi^0\eta n)/\sigma (\pi^0\eta p)$ and
for the $\sigma (\pi^+\eta n)/\sigma (\pi^-\eta p)$ ratios, and partly for the others.

\section{Results}
\label{sec:results}

\begin{figure*}[thb]
\centerline{\resizebox{0.9\textwidth}{!}{%
  \epsfig{file=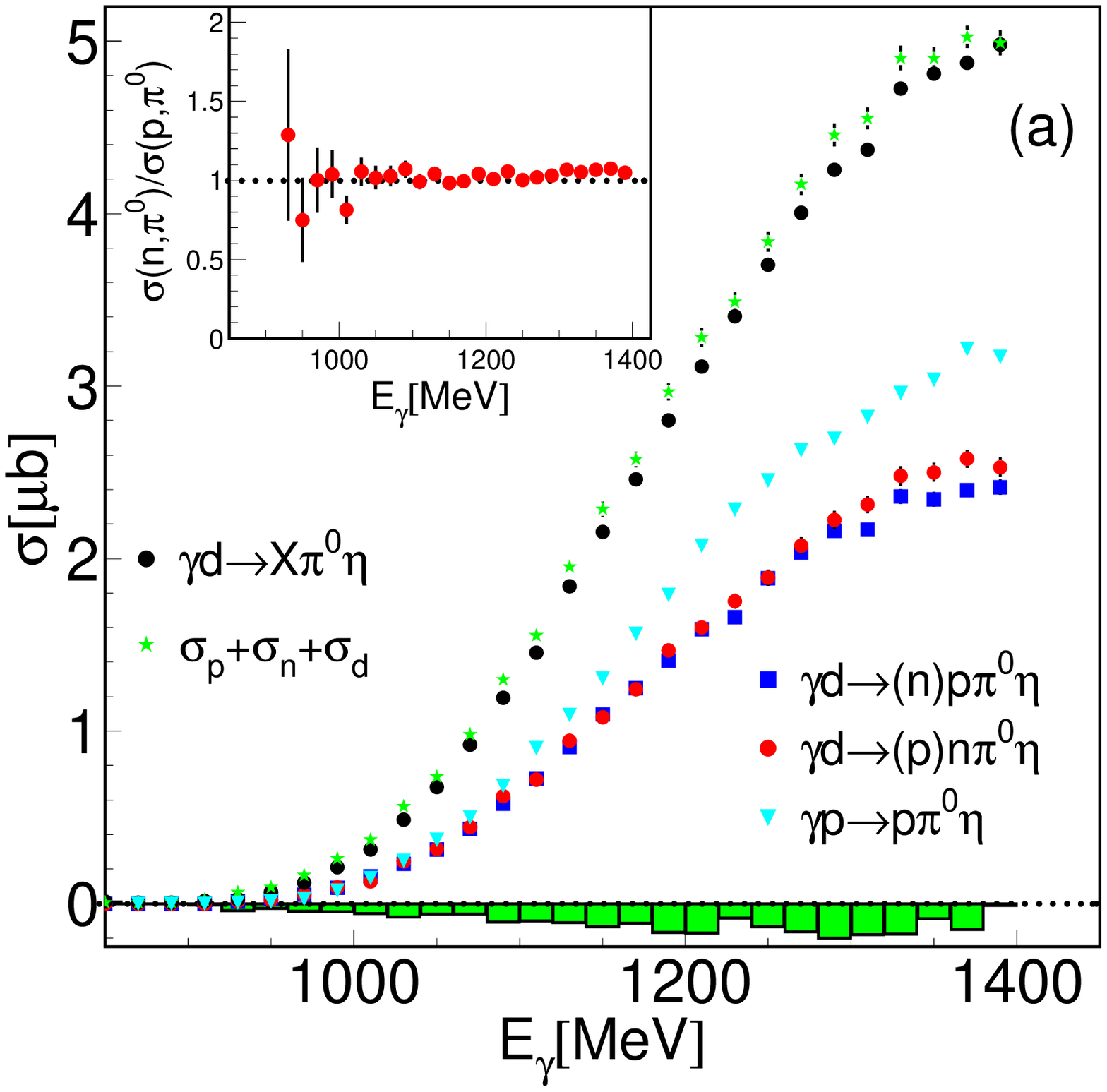}
\hspace*{2cm}\epsfig{file=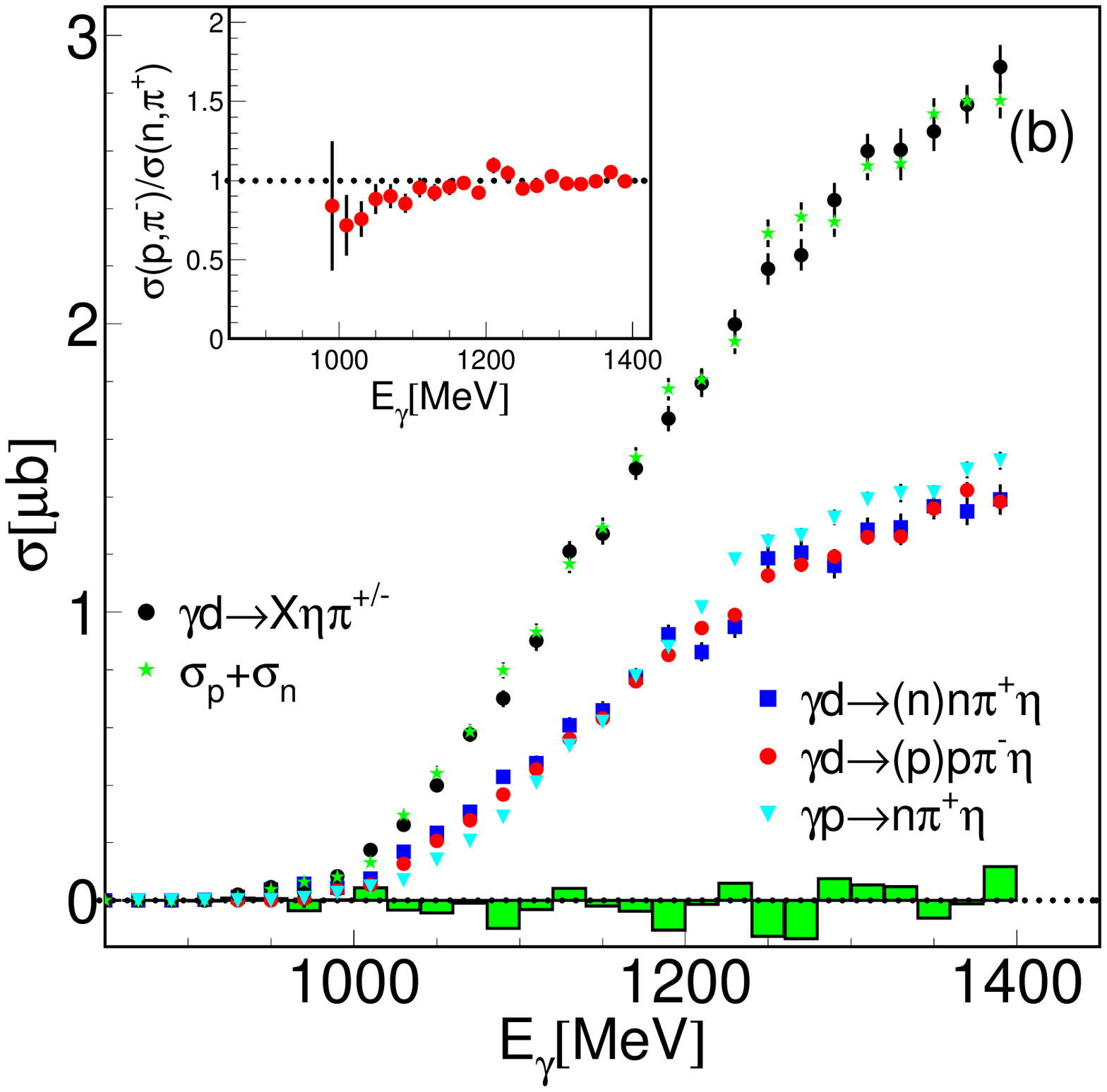}}}
\centerline{\resizebox{0.9\textwidth}{!}{%
  \epsfig{file=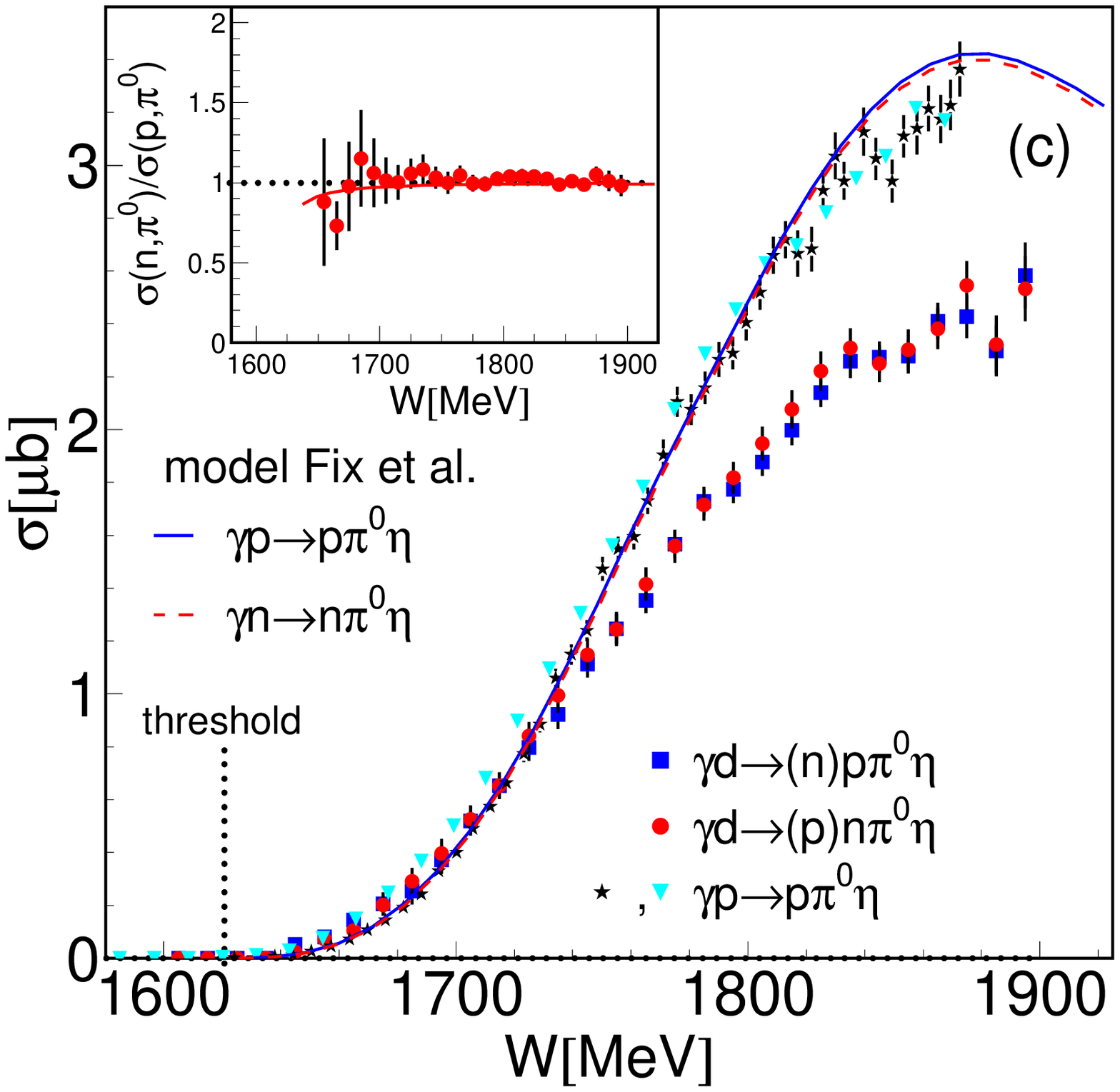}
\hspace*{2cm}\epsfig{file=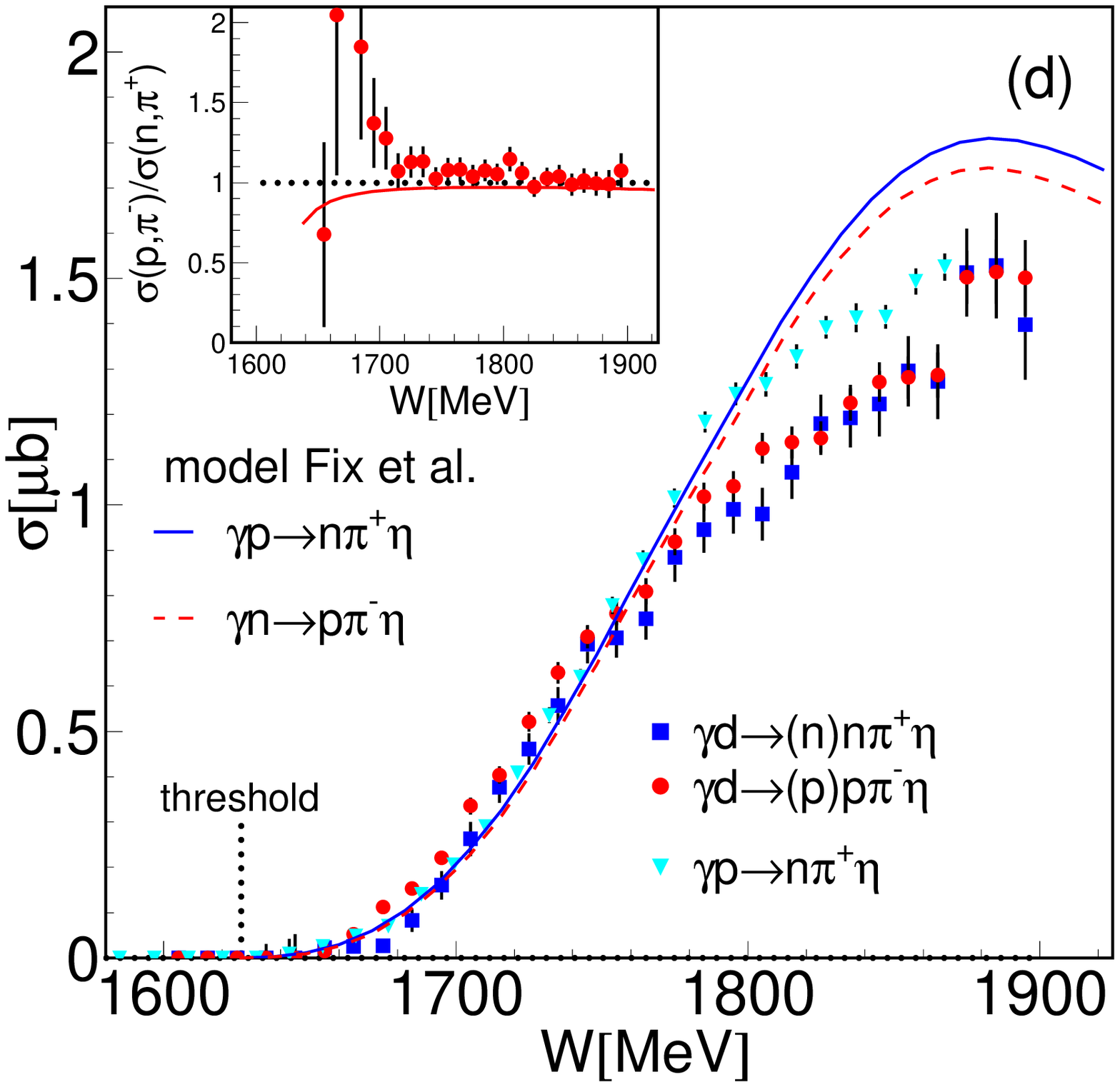}    
}}
\caption{Upper row (a): Total cross sections for the 
$\pi^0\eta N$ final states as a function of incident photon energies. 
(Blue) squares: $p\pi^0\eta$ final state, (red) dots: $n\pi^0\eta$ 
final state, (black) dots: inclusive $X\pi^0\eta$ analysis, (green) stars: 
sum of exclusive cross sections. (Light blue) triangles:  
$p\pi^0\eta$ final state from free protons (hydrogen target). 
Insert: ratio of $\sigma (n,\pi^0)$ to $\sigma (p,\pi^0)$.
(b): Corresponding results for the $\pi^{\pm}\eta N$ final states. 
(Blue) squares: $n\pi^{+}\eta$ final state, (red) dots: $p\pi^{-}\eta$ final 
state, (black) dots: inclusive $X\pi^{\pm}\eta$ analysis, 
(green) stars: sum of exclusive cross sections. 
(Light blue) triangles:  $n\pi^+\eta$ final state from free protons. 
Insert: ratio of $\sigma (p,\pi^{-})$ to $\sigma (n,\pi^{+})$.
For (a) and (b) the green bar histograms indicate the difference between inclusive 
and summed exclusive cross sections.
Bottom row (c): Total cross sections for the $\pi^0\eta N$ final states as a function 
of reconstructed final state invariant mass $W(N_p,\pi^0,\eta)$. 
Notation for results from this work are the same as for (a); 
(black) stars: free proton data from Ref. \cite{Kashevarov_09}.
Model curves from \cite{Fix_10}: solid (blue) free proton, dashed (red)
free neutron target, 
(d): Corresponding results for the $\pi^{\pm}\eta N$ final states. Notation for 
data are the same as in (b), for model curves \cite{Fix_10} as in (c). 
}
\label{fig:total}       
\end{figure*}

\begin{figure*}[thb]
\centerline{\resizebox{0.9\textwidth}{!}{%
\includegraphics{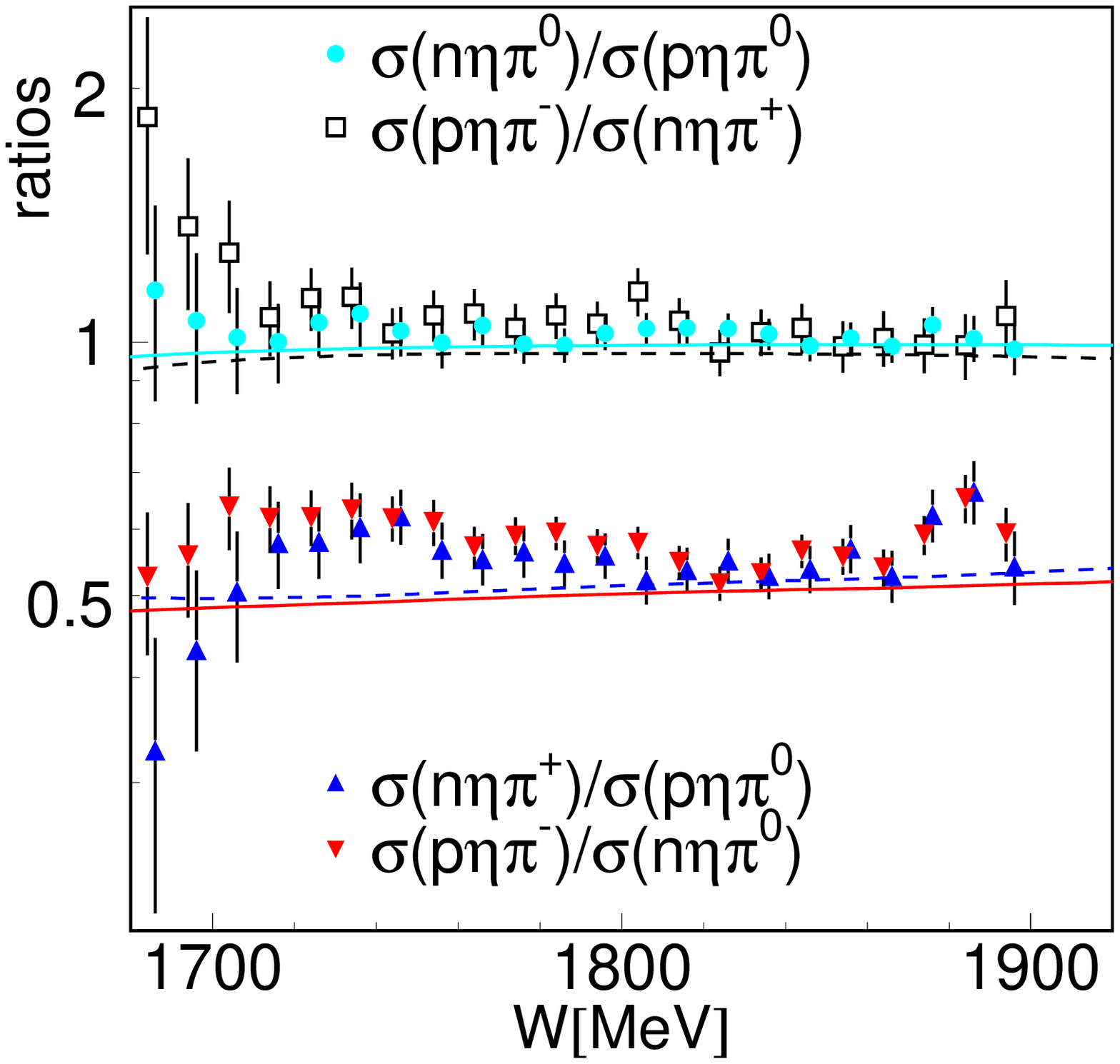}
\hspace*{2cm}\includegraphics{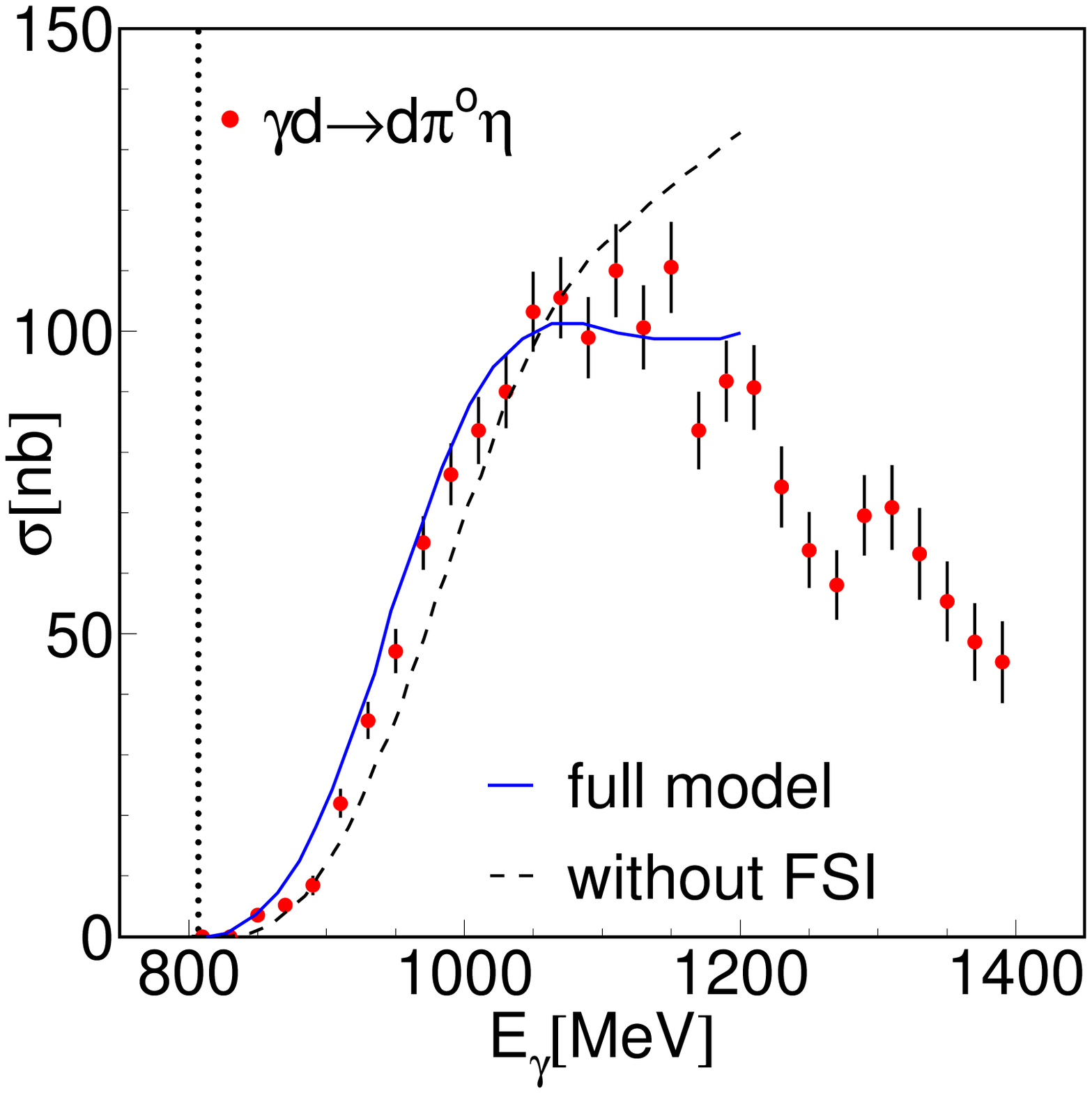}  
}}
\caption{(a) Left hand side: Ratios of total cross sections for the different
isospin channels as a function of final state invariant mass $W(N_p,\pi,\eta)$. 
(b) Right hand side: Total cross section for the coherent 
$\gamma d\rightarrow d\pi^0\eta$ reaction. Model predictions from \cite{Egorov_13}. 
}
\label{fig:ratios}       
\end{figure*}

The results for the total cross sections are summarized in Fig.~\ref{fig:total}.
Figs.~\ref{fig:total}a, \ref{fig:total}b correspond to the analysis as a function of incident 
photon energies, Figs.~\ref{fig:total}c and \ref{fig:total}d show the cross sections extracted 
as a function of reconstructed final state invariant masses. For both types of $\pi\eta$ pairs 
inclusive cross sections and the summed up exclusive cross sections agree with each other
within approximately 5\% (this test is only possible for the analysis using incident photon 
energies since the final state reconstruction requires detection of the recoil nucleons). 
Deviations for the charged pions are statistically distributed but the deviations
for the neutral pions seem to be systematic. A possible systematic effect could arise in the 
inclusive $\pi^0\eta$ cross section because the event generator used for this reaction
did not take into account that events from the coherent reaction have different
angular distributions.

Free-proton data from the measurement with the liquid hydrogen target have also been
analyzed. In the case of $\pi^0\eta$ pairs, the results are in excellent
agreement with previous measurements \cite{Kashevarov_09} (see Fig.~\ref{fig:total}c);
the $\gamma p\rightarrow n\pi^+\eta$ reaction has been measured for the first time.
The results for the $p\pi^0\eta$ final state are in good agreement with the isobar model
of Fix et al. \cite{Fix_10} (which is expected because the model was fitted to the
existing $p\pi^0\eta$ data), but for the $n\pi^+\eta$ final state agreement is not
so good above $W=1.8$~GeV. In contrast to single $\eta$ production \cite{Werthmueller_14}, 
but similar to single $\pi^0$ production \cite{Dieterle_14}, there is a significant 
cross section difference between free and quasi-free results, indicating that  
final state interaction (FSI) effects are present. FSI seems to be larger for 
neutral pions than for charged pions (the same is true for single pion production 
\cite{Krusche_03}). Nevertheless, under the assumption that FSI effects are similar for 
protons and neutrons, cross section ratios can be extracted. They are shown in 
Fig.~\ref{fig:ratios}a. They are in quite good agreement with Eq.~\ref{eq:isorel} 
and also with the model results from \cite{Fix_10}. The ratios for reactions with charged 
and neutral pions are influenced by the different FSI effects. They are in fact somewhat 
larger than predicted. The results for the coherent reaction $\gamma d\rightarrow\pi^0\eta$ 
are compared in Fig.~\ref{fig:ratios}b to model predictions from Egorov and Fix 
\cite{Egorov_13} and good agreement is found here. In summary, the isospin dependence of the 
total cross sections is in excellent agreement with the expectations for a dominant
$\Delta 3/2^-\rightarrow \eta\Delta\mbox{(1232)}\rightarrow \pi\eta N$ contribution.

Differential spectra for invariant mass distributions of the $\pi$-$\eta$ pairs and the 
meson - nucleon pairs have been constructed from full kinematic reconstruction of the final 
states so that they are not effected by Fermi motion. The invariant mass distributions for 
reactions of protons and neutrons and for neutral and charged pions are basically identical
(after re-normalization of their absolute scales) and in excellent agreement with model 
predictions from \cite{Fix_10}. The most prominent feature of this distributions is a 
pronounced structure in the nucleon - pion invariant mass distributions at the invariant mass
of the $\Delta$(1232) resonance in agreement with the assumption of an intermediate
$\eta\Delta$ state in the  
$\Delta 3/2^-\rightarrow \eta\Delta\mbox{(1232)}\rightarrow \pi\eta N$ reaction chain.

Angular distributions have been analyzed as in \cite{Kashevarov_09} in two 
different cm frames, the canonical and the helicity system. Also the angular distributions 
do not show much difference between proton and neutron targets or between neutral and charged 
pions. Agreement with the model predictions from \cite{Fix_10} is reasonable for all observables, 
which is further evidence that the dominating production mechanisms are well understood.

The results for this differential cross sections and also for polarization observables like the
beam-helicity asymmetry measured with a circularly polarized photon beam and an unpolarized target
will be discussed in a detailed paper following the present letter. 

\section{Summary and Conclusions}
Precise results have been obtained for the photoproduction of $\pi^0\eta$ pairs from
free protons, quasi-free protons, quasi-free neutrons bound in deuterium, and
coherently from deuterons. Production of $\pi^+\eta$ pairs has been studied for free 
and quasi-free protons and production of $\pi^-\eta$ pairs for quasi-free neutrons. 
This means that all possible isospin channels for photoproduction of $\pi\eta$-pairs 
from nucleons and from the deuteron have now been measured (previous experiments
covered only the free $\gamma p\rightarrow p\pi^0\eta$ reaction). 
The main results can be summarized as follows: 

The comparison of free and quasi-free measurements for proton targets indicates
significant FSI effects. These effects are quantitatively not precisely understood, 
but seem to be more important for the comparison of reactions with neutral and 
charged pions than for the comparison of final states with the same type of pions 
but different nucleons. This is similar to the photoproduction of single pions 
which shows also much larger effects for neutral pions due to the different FSI 
effects in the $np$ system (which can be bound in the final state) compared to 
the $nn$ and $pp$ systems \cite{Krusche_03}. 
 
The most important results are the cross section ratios for the different isospin
channels of the quasi-free production reactions. They allow model independent conclusions 
about the dominant reaction mechanism. Equation~(\ref{eq:isorel}) is valid for 
(I) $\gamma N\rightarrow \Delta^{\star}\rightarrow \Delta\eta\rightarrow N\pi\eta$ and
(II) $\gamma N\rightarrow \Delta^{\star}\rightarrow N^{\star}\pi\rightarrow N\pi\eta$ 
reaction chains, but not for 
$\gamma N\rightarrow N^{\star}\rightarrow N^{\star}\eta\rightarrow N\pi\eta$ or
$\gamma N\rightarrow N^{\star}\rightarrow N^{\star}\pi\rightarrow N\pi\eta$ sequences
(for the latter $\sigma(\pi^{\pm}\eta N)/\sigma(\pi^0\eta N)$=2 would hold and
the cross section ratios for neutron and proton targets would depend on the corresponding
ratios of the helicity couplings of the primary $N^{\star}$ resonance).
This is clear evidence for a dominant contribution from primary excitation of
$\Delta^{\star}$ resonances. 

After re-normalization of the absolute magnitudes, invariant mass distributions of the 
$\pi$-$\eta$ pairs and the meson - nucleon pairs are in agreement for all four reaction 
channels. They show the same features already observed for the free 
$\gamma p\rightarrow \pi^0\eta p$ reaction \cite{Horn_08a,Horn_08b,Kashevarov_09}).
This is in particular a dominant signal from the decay of the $\Delta(1232)$ intermediate state 
in the $N\pi$ invariant mass. Contributions from the $a_0$ meson in the $\pi\eta$
invariant mass become important only at higher incident photon energies
\cite{Horn_08a,Horn_08b,Kashevarov_09}), and the signal from the N(1535)$1/2^-$ intermediate
state in the $N\eta$ invariant mass is small.  

The experimental results for total cross sections and invariant mass distributions
are in good agreement with model predictions from Fix et al. \cite{Fix_10} apart from
the scale difference for quasi-free reactions due to FSI effects. Also, the angular
distributions are in reasonable agreement with expectations. The good agreement
between the total cross section for the coherent $\gamma d\rightarrow d\pi^0\eta$ 
process and the model predictions from Egorov and Fix \cite{Egorov_13} is further 
evidence that the isospin decomposition of this reaction is well understood. Altogether,
all results support the dominance of the 
$\Delta 3/2^-(1700)\rightarrow \eta\Delta\mbox{(1232)}\rightarrow \pi^0\eta p$ reaction chain
in the threshold region.

\vspace*{1cm}
{\bf Acknowledgments}

We wish to acknowledge the outstanding support of the accelerator group 
and operators of MAMI. 
This work was supported by Schweizerischer Nationalfonds
(200020-132799,121781,117601,113511), Deutsche
For\-schungs\-ge\-mein\-schaft (SFB 443, SFB/TR 16, SFB1044), DFG-RFBR (Grant No. 05-02-04014),
UK Science and Technology Facilities Council, (STFC 57071/1, 50727/1), 
European Community-Re\-search Infrastructure Activity (FP6), the US DOE, US NSF and
NSERC (Canada). A. Fix acknowledges support from the Dynasty Foundation, the TPU Grant 
LRU-FTI-123-2014, and the MSE program Nauka (Project 3.825.2014/K).
We thank the undergraduate students of Mount Allison University and The George Washington  
University for their assistance.

\end{document}